\documentclass[preprint,12pt]{article}
\usepackage{amssymb}
\usepackage{amsmath}

\usepackage{graphicx}
\usepackage{float} 
\usepackage[colorlinks=true, allcolors=blue]{hyperref} 

\usepackage{caption}
\begin{document}


\title{Physics-Informed Neural Networks for Irregular Domain Mapping and Partial Differential Equations solving} 

\author{Cuizhi Zhou, Kaien Zhu$^{*}$\\
\small{School of Physics, Nankai University, Tianjin 300071, People's Republic of China}\\
\small{*Email: zhukaien@nankai.edu.cn}}
\date{\today}
\maketitle

\begin{abstract}

The solution of partial differential equations (PDES) on irregular domains has long been a subject of significant research interest.  In this work, we present an approach utilizing physics-informed neural networks (PINNs) to achieve irregular-to-regular domain mapping. 
Thus we can use finite difference method and physics-informed convolutional neural networks to solve PDEs on rectangular grids instead of the original irregular boundary. 

Structured grids on irregular domains are obtained by inverse mapping. 
We demonstrate PINN's versatile capability to produce customized structured grids tailored to diverse computational requirements, thereby significantly facilitating PDEs solving.

\end{abstract}

Structured grids, Partial Differential Equations 


\section{Introduction}
\label{sec1}

The numerical solution of partial differential equations (PDEs) on irregular domains has long been a challenging problem. Over the past few decades, the primary numerical methods for addressing this issue have been the Finite Element Method (FEM)\cite{astuto2024nodal} and the Finite Volume Method (FVM) \cite{HOLLBACHER2019186}. However, these methods impose stringent requirements on mesh quality. In 2019, M. Raissi et al. proposed the Physics-Informed Neural Network (PINN) model \cite{RAISSI2019686}, which integrates the residuals of PDE governing equations, boundary conditions, and initial conditions into a multi-objective loss function. This approach enables the network to inherently satisfy physical laws during training. By eliminating the reliance on computational grids, PINN achieves direct modeling in continuous space, offering a novel solution for solving PDEs on irregular domains.

Despite its widespread application in PDE solving, PINN still suffers from low accuracy and excessively long training times when applied to irregular domains. Several improvements have been proposed to address these limitations. Complex geometries are embedded into a regular background grid, with the loss function distinguishing between interior and exterior domains to avoid explicit mesh generation \cite{KHARA2024103709,Peskin_2002}. Distance functions are introduced into the loss function to weight residuals in the domain and on the boundary \cite{SUKUMAR2022114333,McFall20091221,SHENG2021110085}. Dual-grid approaches \cite{lu2021learning}  use one network satisfies the PDE and the other one 
 handles boundary conditions . Domain mapping is presented in \cite{GAO2021110079} where  irregular domains are mapped to regular ones. 
 We are interested in the domain mapping approach in \cite{GAO2021110079} where
elliptic coordinates are used for domain mapping and physics-informed convolutional neural networks (PICNN) is used for PDEs solving. 

In this article we map irregular boundaries of PDEs to regular domains by PINN.  We will show that PINN can generate customized grids.  Structured meshes are obtained by inverse mapping. We can solve PDE on irregular domain with FEM or FVM on structured grid, or in a better way, we can solve mapped PDE on rectangular grids with  finite difference method (FDM) or spectral method.

The paper is organized as follows: Section II Solve the PDE Equation in the Computational Space; Section III introduce grids generation techniques.

\section{Solving PDEs  in the Computational Space}

Solving PDEs on irregular domains has long been a challenging problem. The primary numerical methods for addressing this issue have been the FEM\cite{astuto2024nodal} and the FVM \cite{HOLLBACHER2019186}.  Can we map irregular boundaries to rectangular boundaries so that FDM or spectral method can work?  

This section introduces a method to construct the mapping from irregular domains to regular domains using PINNs, thereby enabling efficient PDE solver such as FDM in the computational space (regular domain) . The rectangular space is referred to as the computational space, and the original space is called the physical space.

\subsection{Domain mapping and PINN architecture}

\begin{figure}[htbp]
    \centering
    \includegraphics[width=0.5\textwidth]{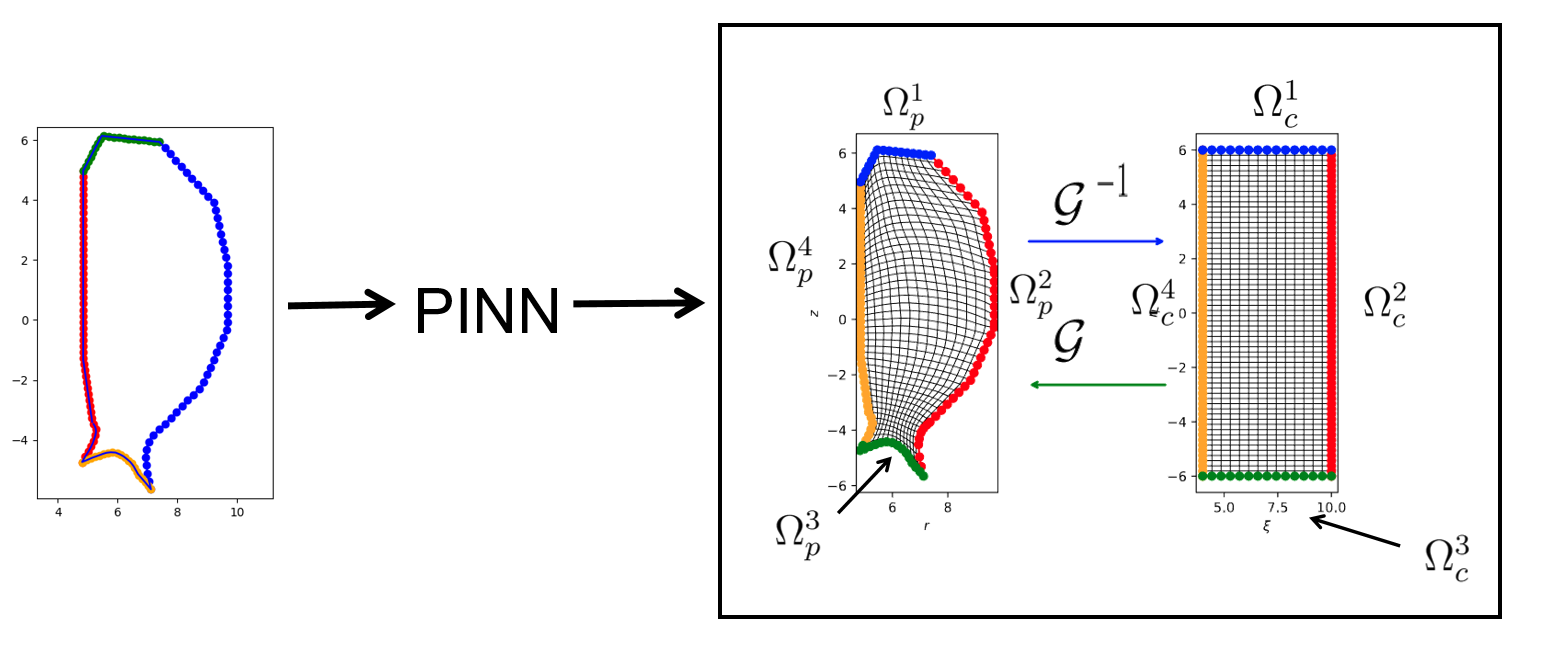}
    \caption{The PINN mapping approach.}
    \label{fig2}
\end{figure}

 We follow the notation system in  reference \cite{GAO2021110079}.  Grids are obtained by PINN method.  The key point is to find the regular-to-irregular domain mapping $\mathcal{G}: \Omega_p \rightarrow \Omega_c$ in Fig. \ref{fig2} with which structured meshes are obtained,
\begin{equation}
    \mathbf{x} = \mathcal{G}(\boldsymbol{\xi}), \quad  \forall \xi \in \partial \Omega_{c}^i, \, i = 1, \ldots, 4,
    \label{eq:mapping1}
\end{equation}
\begin{equation}
    \quad \boldsymbol{\xi} = \mathcal{G}^{-1}(\mathbf{x}), \quad \forall x \in \partial \Omega_{p}^i, \, i = 1, \ldots, 4.
     \label{eq:mapping2}
\end{equation}

PINN employs a Multi - Layer Perceptron (MLP) architecture in this article. 
The input of this neural network consists of the coordinates $(\xi,\eta)$ in the computational space, while its output is the coordinates $(x,y)$ in the physical space.

\begin{figure}[htbp]
    \centering
    \includegraphics[width=\textwidth]{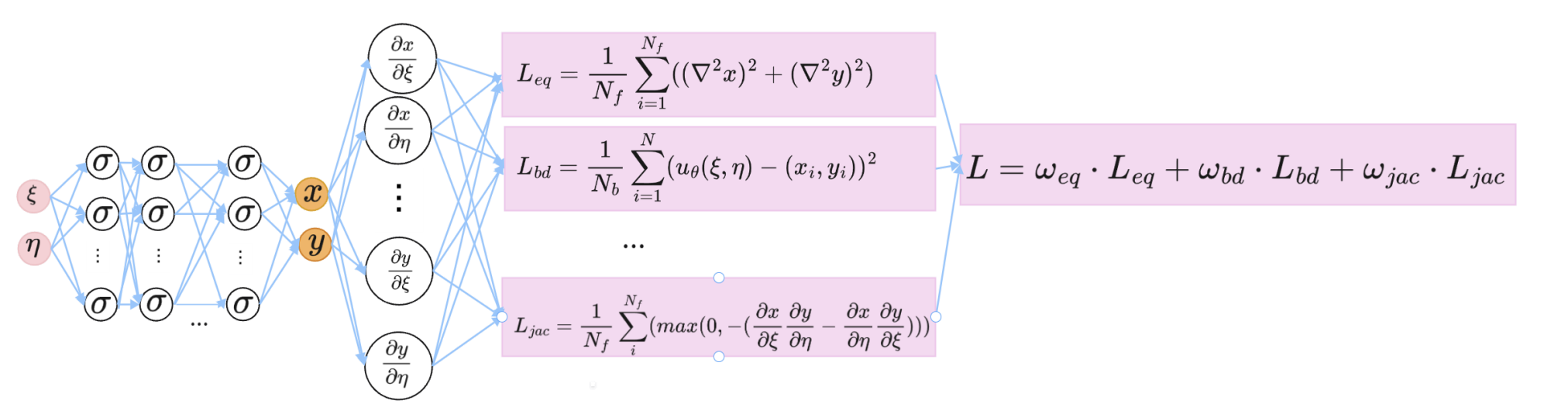}
    \caption{Schematic diagram of the PINN model.}
    \label{network}
\end{figure}

The loss function of PINN is:

\begin{equation}
    L = \omega_{eq} \cdot L_{eq} + \omega_{bd} \cdot L_{bd} + \omega_{jac} \cdot L_{jac},
    \label{eq:total_loss}
\end{equation}
where
\begin{equation}
L_{eq}=\frac{1}{N_f}\sum_{i = 1}^{N_f}((\nabla^2 x)^2 +(\nabla^2 y) ^2) ,
\label{eq:smooth_eq}
\end{equation}
\begin{equation}
L_{bd}=\frac{1}{N_b}\sum_{i = 1}^{N}(u_\theta(\xi,\eta)-(x_i,y_i))^2,
\end{equation}
\begin{equation}
    L_{jac} = \frac{1}{N_f}\sum_{i}^{N_f} (max(0,-(\frac{\partial x}{\partial \xi}\frac{\partial y}{\partial \eta} - \frac{\partial x}{\partial \eta} \frac{\partial y}{\partial \xi}))),
\end{equation}
\begin{equation}
    \nabla^2 x = \frac{\partial^2 x}{\partial \xi^2} + \frac{\partial^2 x}{\partial \eta^2}  , \quad \forall (\xi,\eta) \in \Omega_c,
    \label{eq: ellipse_simple1}
\end{equation}
\begin{equation}
 \nabla^2 y = \frac{\partial^2 y}{\partial \xi^2} + \frac{\partial^2 y}{\partial \eta^2} , \quad \forall (\xi,\eta) \in \Omega_c.
  \label{eq: ellipse_simple2}
\end{equation}

The term $L_{eq}$ means solving a diffusion equation in computational space which is different from  \cite{GAO2021110079} where a diffusion equation is solved in physical space.  Laplacian-based regularization term of the loss function  enforces local smoothness of the grid distribution. 

The boundary loss function is $L_{bd}$, where $N_b$ denotes the total number of discretely sampled points on the boundary, $u_\theta$ represents the mapping function constructed by the neural network with parameters $\theta$, and $(x_i,y_i)$ corresponds to the precise coordinate values of the $i$-th sampling point on the boundary of the physical computational domain.

To ensure the generated grid cells maintain valid topological structures, this paper introduces a Jacobian determinant constraint as a key regularization term $L_{jac}$,  whose core objective is to enforce the mathematical condition $det(J) > 0 $ for each grid cell's Jacobian matrix.

\begin{figure}
    \centering
    \includegraphics[width=0.5\linewidth]{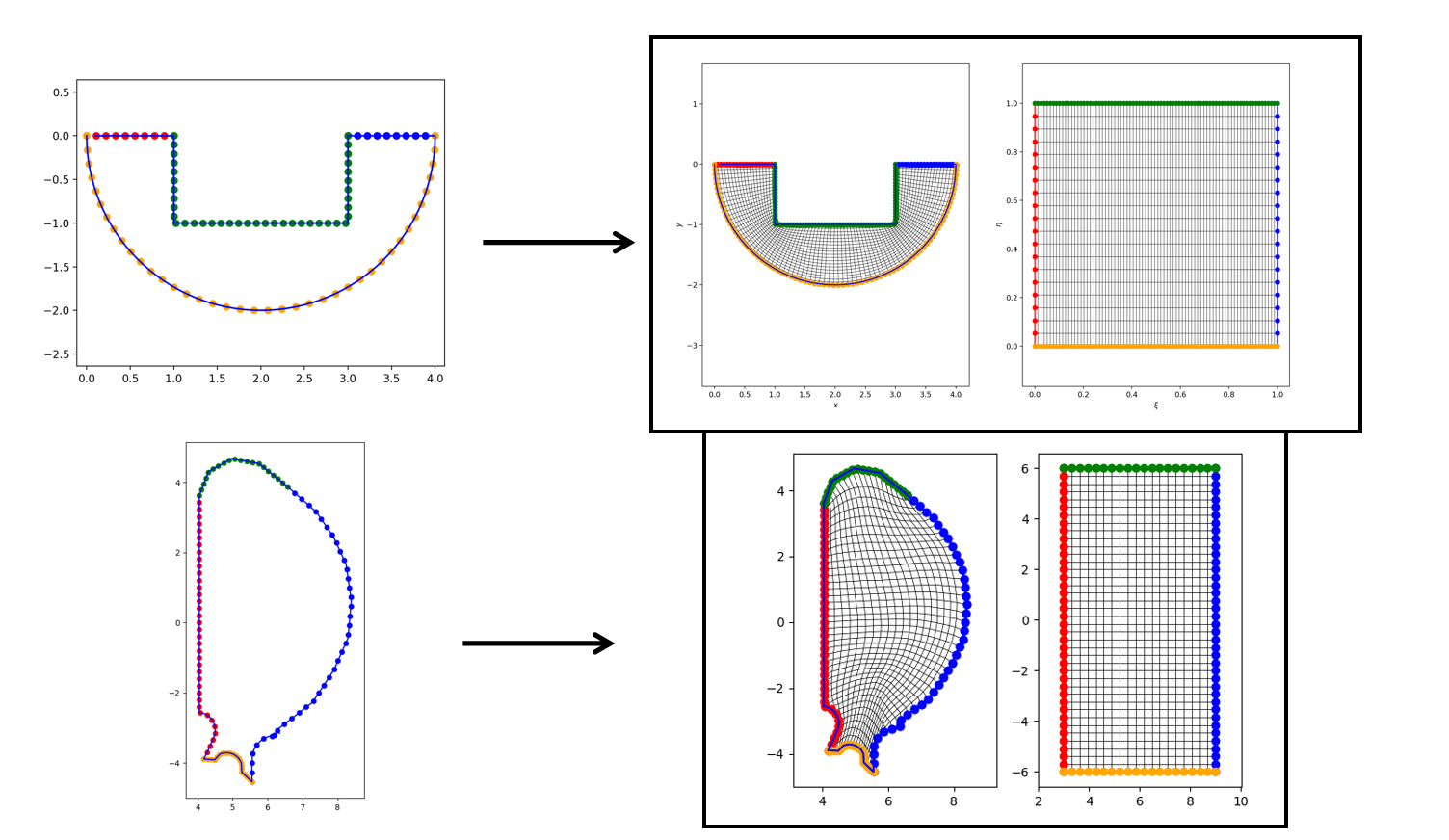}
    \caption{ PINN-generated  meshes on irregular domains.}
    \label{fig:example}
\end{figure}

By conducting optimization training based on the above composite loss function $L$, grids with excellent characteristics can be obtained. Figure \ref{fig:example} illustrates some numerical results. The mesh generated by PINN approach has high quality standards.

\subsection{The technique to solve PDEs with irregular boundary }
Irregular boundaries are mapped to rectangles and it is easy to get the grids of rectangles, which is shown in Fig \ref{fig:example}. But how to use rectangular mesh to solve PDEs? The answer is to map equations of physical space to the ones of computational space since the map  $\mathcal{G}$ and $\mathcal{G}^{-1}$ are established by PINN. 

\subsubsection{Transformation of equations to computational space}

The transformation framework from physical space to computational space could be found in \cite{anderson1995computational}.  For ease of reference, the key points of the framework  are presented here. 

The transformation relationship of differential operators between the physical space and the computational space can be characterized by the following mathematical form:

\begin{equation}
\frac{\partial}{\partial x} = \frac{\partial}{\partial \xi} \frac{\partial \xi}{\partial x} + \frac{\partial}{\partial \eta} \frac{\partial \eta}{\partial x},
\label{eq:partial_x_compact}
\end{equation}

\begin{equation}
\frac{\partial}{\partial y} = \frac{\partial}{\partial \xi} \frac{\partial \xi}{\partial y} + \frac{\partial}{\partial \eta} \frac{\partial \eta}{\partial y}.
\label{eq:partial_y_compact}
\end{equation}

The second-order derivatives can be derived:

\begin{equation}
\begin{split}
\frac{\partial^2}{\partial x^2} &= \frac{\partial}{\partial x} \left( \frac{\partial}{\partial x} \right) \\
&= \frac{\partial}{\partial x} \left( \frac{\partial \xi}{\partial x} \frac{\partial}{\partial \xi} + \frac{\partial \eta}{\partial x} \frac{\partial}{\partial \eta} \right) \\
&= \left( \frac{\partial \xi}{\partial x} \frac{\partial}{\partial \xi} + \frac{\partial \eta}{\partial x} \frac{\partial}{\partial \eta} \right) \left( \frac{\partial \xi}{\partial x} \frac{\partial}{\partial \xi} + \frac{\partial \eta}{\partial x} \frac{\partial}{\partial \eta} \right) \\
&= \left( \frac{\partial \xi}{\partial x} \right)^2 \frac{\partial^2}{\partial \xi^2} + 2 \frac{\partial \xi}{\partial x} \frac{\partial \eta}{\partial x} \frac{\partial^2}{\partial \xi \partial \eta} \\
&\quad + \left( \frac{\partial \eta}{\partial x} \right)^2 \frac{\partial^2}{\partial \eta^2} + \frac{\partial^2 \xi}{\partial x^2} \frac{\partial}{\partial \xi} + \frac{\partial^2 \eta}{\partial x^2} \frac{\partial}{\partial \eta},
\end{split}
\end{equation}
\begin{equation}
\begin{split}
\frac{\partial^2}{\partial y^2} &= \frac{\partial}{\partial y} \left( \frac{\partial}{\partial y} \right) \\
&= \frac{\partial}{\partial y} \left( \frac{\partial \xi}{\partial y} \frac{\partial}{\partial \xi} + \frac{\partial \eta}{\partial y} \frac{\partial}{\partial \eta} \right) \\
&= \left( \frac{\partial \xi}{\partial y} \frac{\partial}{\partial \xi} + \frac{\partial \eta}{\partial y} \frac{\partial}{\partial \eta} \right) \left( \frac{\partial \xi}{\partial y} \frac{\partial}{\partial \xi} + \frac{\partial \eta}{\partial y} \frac{\partial}{\partial \eta} \right) \\
&= \left( \frac{\partial \xi}{\partial y} \right)^2 \frac{\partial^2}{\partial \xi^2} + 2 \frac{\partial \xi}{\partial y} \frac{\partial \eta}{\partial y} \frac{\partial^2}{\partial \xi \partial \eta} \\
&\quad + \left( \frac{\partial \eta}{\partial y} \right)^2 \frac{\partial^2}{\partial \eta^2} + \frac{\partial^2 \xi}{\partial y^2} \frac{\partial}{\partial \xi} + \frac{\partial^2 \eta}{\partial y^2} \frac{\partial}{\partial \eta}.
\end{split}
\end{equation}
Laplace operator is
\begin{equation}
    \nabla^2 T  = \frac{1}{|J|} \left( \frac{\partial}{\partial \xi} \left( |J| \frac{\partial T  }{\partial \xi} \right) + \frac{\partial}{\partial \eta} \left( |J| \frac{\partial T}{\partial \eta} \right) \right) 
    \label{eq:laplace_operator}
\end{equation}
where $J$ is the Jacobian matrix

\begin{equation}
    J = \begin{bmatrix}
        \dfrac{\partial x}{\partial \xi} & \dfrac{\partial x}{\partial \eta} \\[0.5em]
        \dfrac{\partial y}{\partial \xi} & \dfrac{\partial y}{\partial \eta}
    \end{bmatrix}.
\end{equation}

\subsubsection{FDM on computational domain}

\begin{figure}[H]
    \centering
    \includegraphics[width=0.5\linewidth]{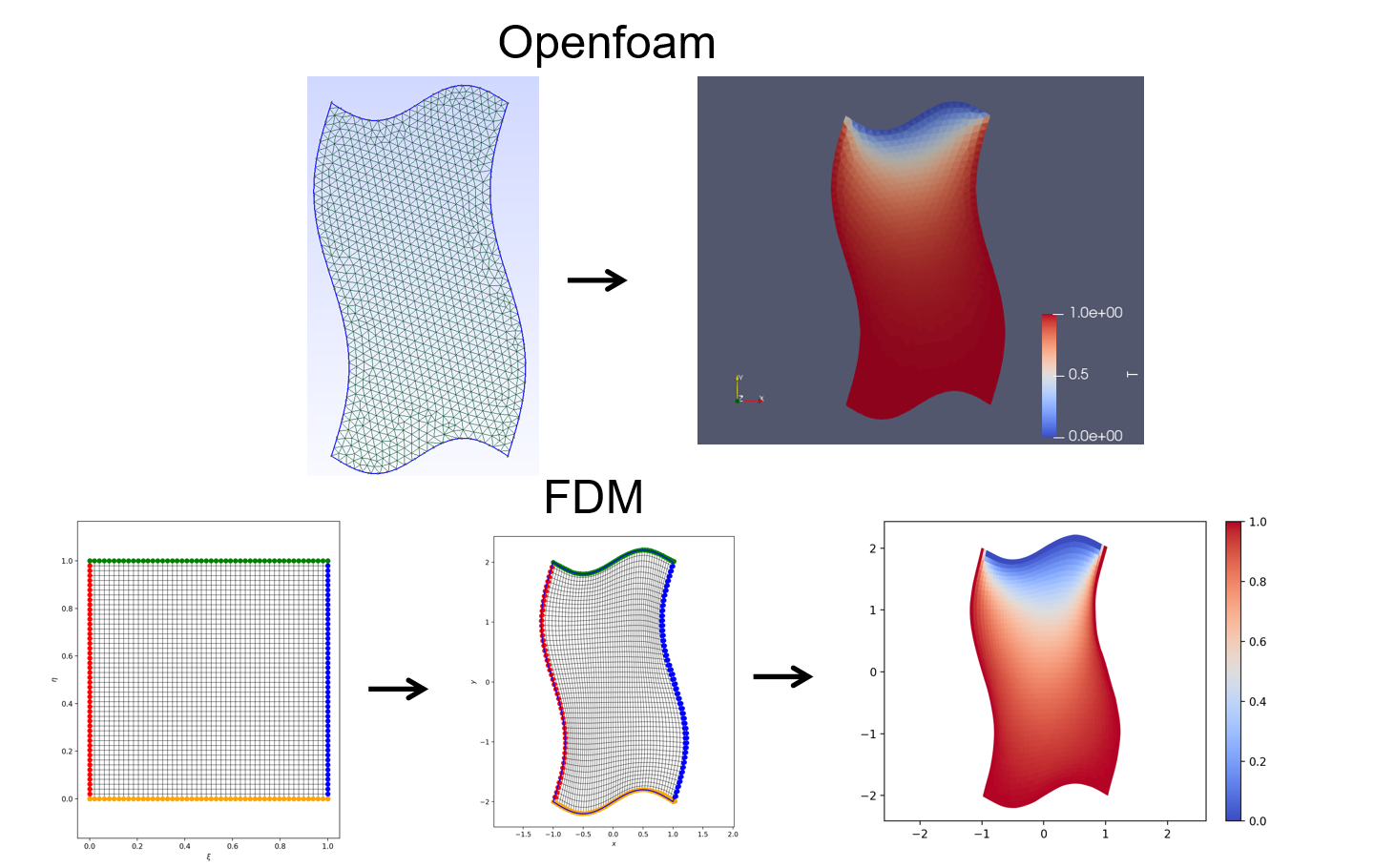}
    \caption{ Temperature distribution by OpenFOAM and FDM.}
    \label{fig:OpenFoam vs. FDM}
\end{figure}

Take heat conduction equation with irregular boundary as an example.  The boundary is shown in Fig. \ref{fig:OpenFoam vs. FDM}.  The equation is
\begin{equation}
    \nabla^2 T=\frac{\partial^2T}{\partial x^2}+\frac{\partial^2 T}{\partial y^2}=0, \quad \mathbf{(x,y)} \in \Omega_p
\end{equation}
\begin{equation}
    T(\mathbf{(x,y)}) = 0, \quad \mathbf{(x,y)} \in \partial \Omega_{\text{p,up}} 
\end{equation}
\begin{equation}
     T(\mathbf{(x,y)}) = 1, \quad \mathbf{(x,y)} \in \partial \Omega_{\text{p, left}}  ,\Omega_{\text{p, right}} ,\Omega_{\text{p, down}} 
\end{equation}
where $T$ is the temperature field of 2D domain. In the computational space, we need to solve the following equation:
\begin{equation}
    \nabla^2 T  = \frac{1}{|J|} \left( \frac{\partial}{\partial \xi} \left( |J| \frac{\partial T  }{\partial \xi} \right) + \frac{\partial}{\partial \eta} \left( |J| \frac{\partial T}{\partial \eta} \right) \right) =0,\quad \mathbf{(x,y)} \in \Omega_c
    \label{eq:heat_equation computational}
\end{equation}
\begin{equation}
    T(\mathbf{(\xi,\eta)}) = 0, \quad \mathbf{(\xi,\eta)} \in \partial \Omega_{\text{c, up}} 
\end{equation}
\begin{equation}
     T(\mathbf{(\xi,\eta)}) = 1, \quad \mathbf{(\xi,\eta)} \in \partial \Omega_{\text{c, left}}  ,\Omega_{\text{c, right}} ,\Omega_{\text{c, down}} 
\end{equation}
FDM are used to solve the above equation and the data are mapped back to physical domain. This approach works quite satisfactory and  the the results are presented in Fig. \ref{fig:OpenFoam vs. FDM}.  The benchmark is also presented in Fig. \ref{fig:OpenFoam vs. FDM} where FVM is used to solve the equation in OpenFOAM software.

\subsection{Solve equation using PICNN}

After the grids are established, there is another way to solve PDEs in computational domain where neural networks are used.  In fact, physics-informed convolutional neural networks (PICNN)  is such a technique \cite{GAO2021110079}. 

Since convolutional neural network (CNN) only can handle regular geometries with image-like format ,  Gao et al.  applied elliptic coordinate mapping to enable coordinate transforms between the irregular physical domain and regular computational domain \cite{GAO2021110079}, that is, they choose first to map 
\begin{equation}
\frac{\partial^2 \xi}{\partial x^2} + \frac{\partial^2 \xi}{\partial y^2} = 0,
\label{eq:ellipse1}
\end{equation}
\begin{equation}
\frac{\partial^2 \eta}{\partial x^2} + \frac{\partial^2 \eta}{\partial y^2} = 0,
\label{eq:ellipse2}
\end{equation}
to computational domain and then to solve the mapped equations by traditional FDM algorithm.  

Our approach in this article is to solve elliptic  equation by PINN in computational domain directly,  that is, Eq. \ref{eq: ellipse_simple1} and Eq. \ref{eq: ellipse_simple2}. 
The equations are simpler and the PINN way is more flexible about which will be presented later.   

\begin{figure}
    \centering
    \includegraphics[width=0.5\linewidth]{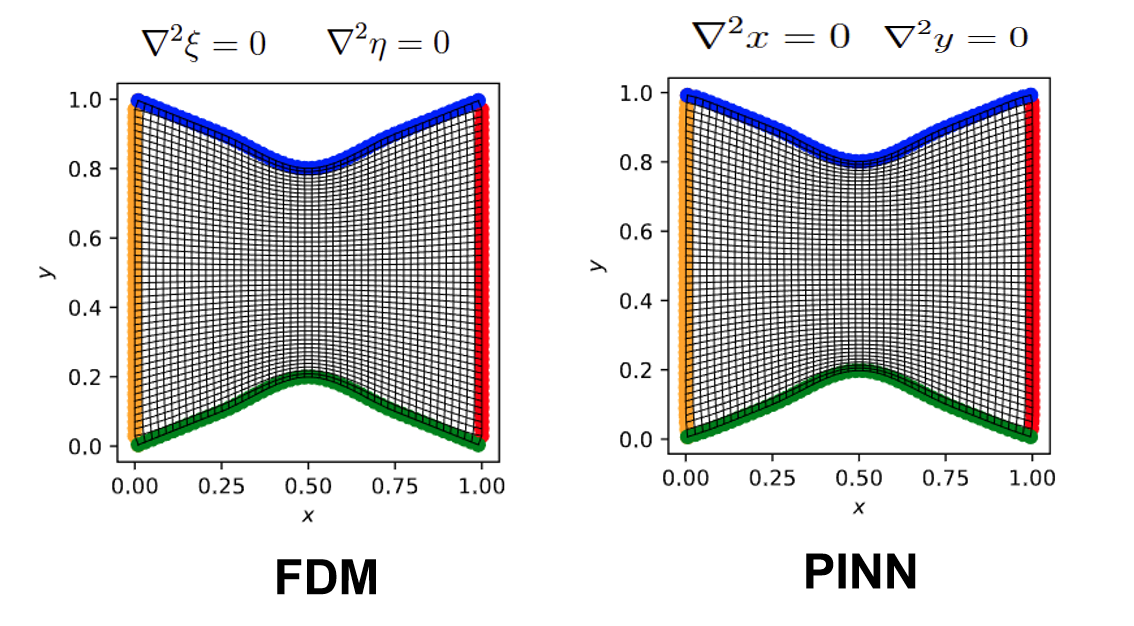}
    \caption{Comparison between our PINN-generated mesh and FDM-generated mesh of \cite{GAO2021110079}}.
    \label{fig:compare_mesh}
\end{figure}

Figure \ref{fig:compare_mesh} presents the meshes generated by our PINN approach and FDM approach of \cite{GAO2021110079}.  Both methods produce high-quality meshes.

The steady Navier-Stokes equations for two-dimensional incompressible fluid problems is
\begin{equation}
\begin{cases}
\begin{aligned}
& \frac{\partial u}{\partial x} + \frac{\partial v}{\partial y} = 0 \\[10pt]
& \frac{\partial u}{\partial t} + u \frac{\partial u}{\partial x} + v \frac{\partial u}{\partial y} = -\frac{1}{\rho} \frac{\partial p}{\partial x} + \nu \left( \frac{\partial^2 u}{\partial x^2} + \frac{\partial^2 u}{\partial y^2} \right) \\[10pt]
& \frac{\partial v}{\partial t} + u \frac{\partial v}{\partial x} + v \frac{\partial v}{\partial y} = -\frac{1}{\rho} \frac{\partial p}{\partial y} + \nu \left( \frac{\partial^2 v}{\partial x^2} + \frac{\partial^2 v}{\partial y^2} \right)
\end{aligned}
\end{cases}
\end{equation}
where $u$ and $v$ represent the components of the velocity in the $x$ and $y$ directions respectively, $p$ represents the pressure field, $\rho$ is the density of the fluid, and $\nu$ is the kinematic viscosity coefficient.   

\begin{figure}
    \centering
    \includegraphics[width=0.5\linewidth]{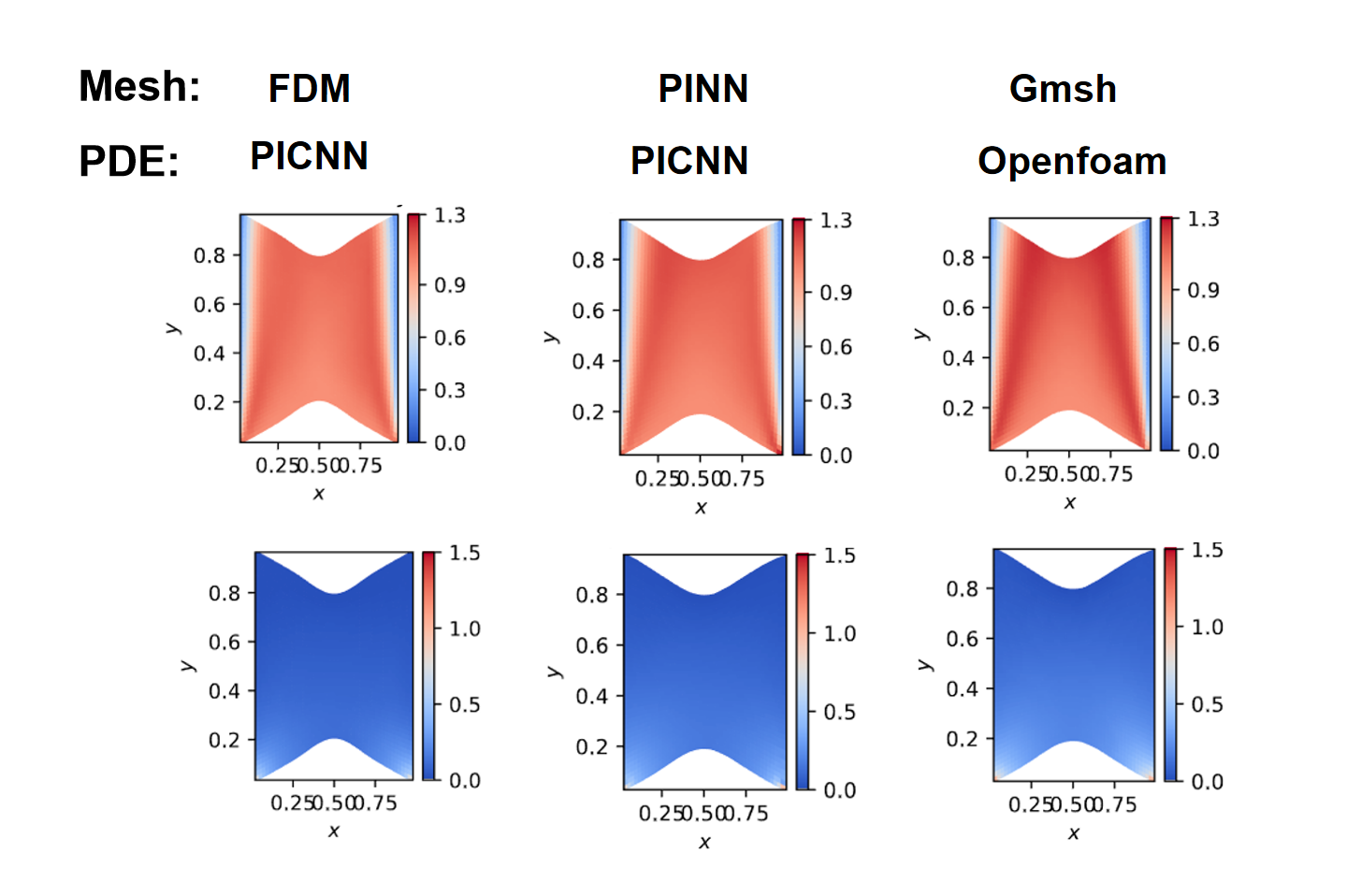}
    \caption{Navier-Stokes equations solved by: (left)  PICNN on FDM-generated grids; (middle) PICNN on PINN-generated grids; (right) OpenFOAM's solution results of FVM.}
    \label{fig:Navier-stokes}
\end{figure}

Figure \ref{fig:Navier-stokes} displays the numerical solutions of the Navier-Stokes equations, where left column shows results obtained using PICNN on FDM-generated meshes (method  of \cite{GAO2021110079}), middle column  presents solutions computed with PICNN on PINN-generated meshes (our approach), and right column provides reference solutions from OpenFOAM's FVM. The comparative results demonstrate that the PICNN on PINN-generated meshes  is a creditable approach for PDEs with irregular boundaries.

\section{Grid generation techniques}

In regions such as boundary layers, shock waves, chemical reaction interfaces, and fluid vortices, the spatial gradients of solution functions are extremely large, meaning that physical quantities change sharply over short distances. Therefore, we need to divide grids densely in such regions. Meanwhile, when the computational domain contains structures such as sharp corners or small holes, it is also necessary to divide grids densely.

Traditional body-fitted coordinate methods typically need to construct and solve sophisticated high-order partial differential equations (e.g., harmonic mapping equations, Winslow equations) to achieve specific grid characteristics.   In this article we will show that PINN-based grid mapping approach is  versatile for such situations  by mathematically formulating requirements as loss functions. 

Different requirements, such as boundary conformity, orthogonality, cell size distribution, etc., could be easily added into the loss function. This framework provides two major advantages: (1) eliminating the dual challenges of equation formulation and numerical solution in conventional methods, and (2) simultaneously handling geometric complexity and physical constraints.  

\subsection{Basic Smooth Grid}

\begin{figure}[H]
    \centering
    \includegraphics[width=0.5\linewidth]{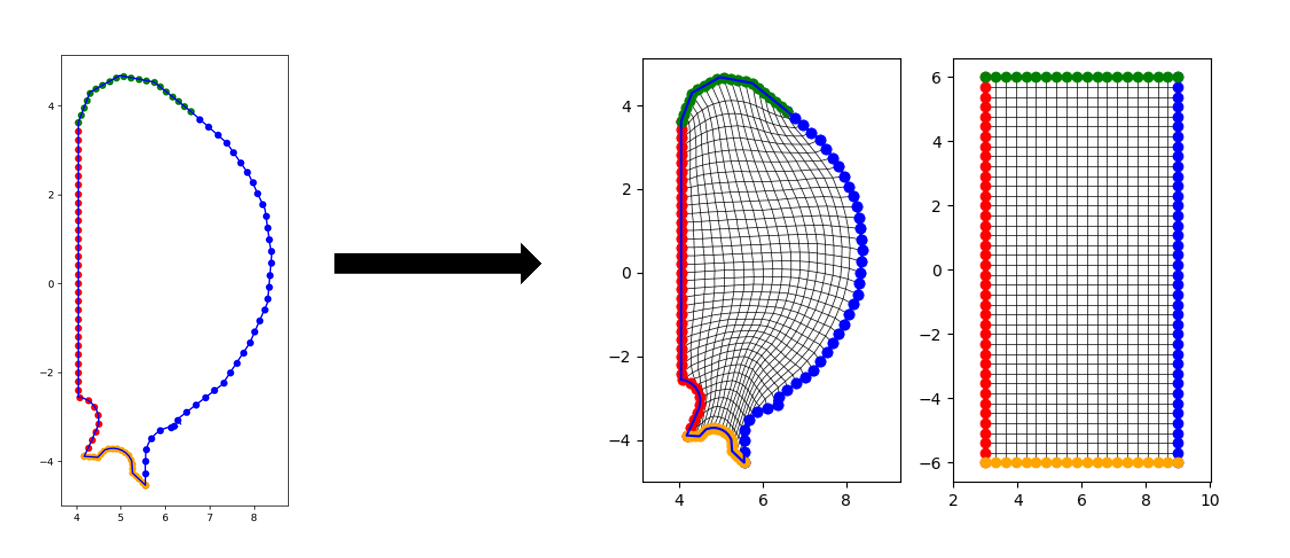}
    \caption{Mesh generation for a tokamak's section.}
    \label{fig:iter}
\end{figure}

Smooth grids are the primary consideration in grid generation.  Equation (\ref{eq:smooth_eq})
 is corresponded for this purpose.  The preceding grids are obtained in this way. 
 Figure \ref{fig:iter} is another example for this basic grid-genereation skill.
 
\subsection{Denser boundary layer grids}

\begin{figure}
    \centering
    \includegraphics[width=0.5\linewidth]{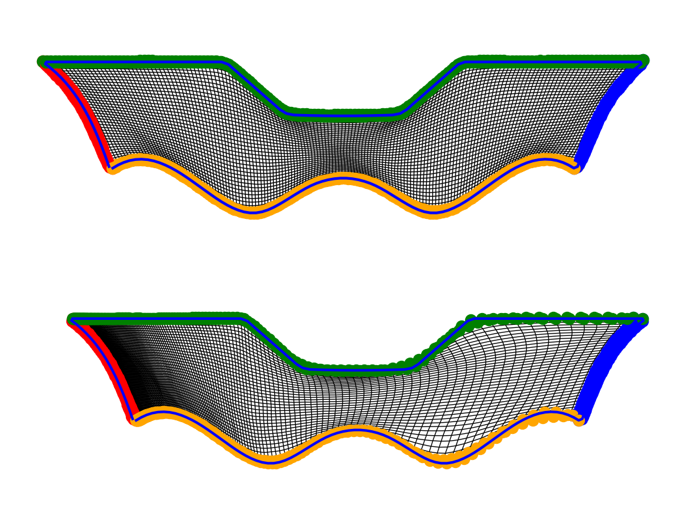}
    \caption{Normal boundary layer (upper) vs. denser boundary layer (lower).}
    \label{fig:dense_boundary_layer}
\end{figure}

Normal basic grids are generated where the boundaries are sampled evenly. 
Take the green boundary line of the upper figure of Fig. \ref{fig:dense_boundary_layer} for example.  Let $t_k (k=0,1,...,N-1)$ be evenly sampled points of  $[0,1]$ and $L$ be the total length of the green line. Let $s_k$ denotes the arc length of the  natural coordinate system. Then 
\begin{equation}
    s_k = L t_k
\end{equation}
is the sampling points of green boundary line. The same procedure is applied to the orange boundary line and we get the grids shown in the upper figure of Fig. \ref{fig:dense_boundary_layer}.

In practical applications of computational grids, some boundary regions often require higher mesh density to enhance calculation accuracy. To address this requirement, this study employs a exponential-function-based non-uniform sampling strategy that mathematically achieves gradient mesh refinement. The formula,

\begin{equation}
    s_k = L \cdot \frac{e^{\alpha t_k} - 1}{e^\alpha - 1},
\end{equation}
gives out denser boundary layer grids, which is shown in the lower figure of Fig. \ref{fig:dense_boundary_layer}. The parameter $\alpha$ controls the density of grids.

\subsection{Sparser and denser grids around some points}

\begin{figure}
    \centering
    \includegraphics[width=0.5\linewidth]{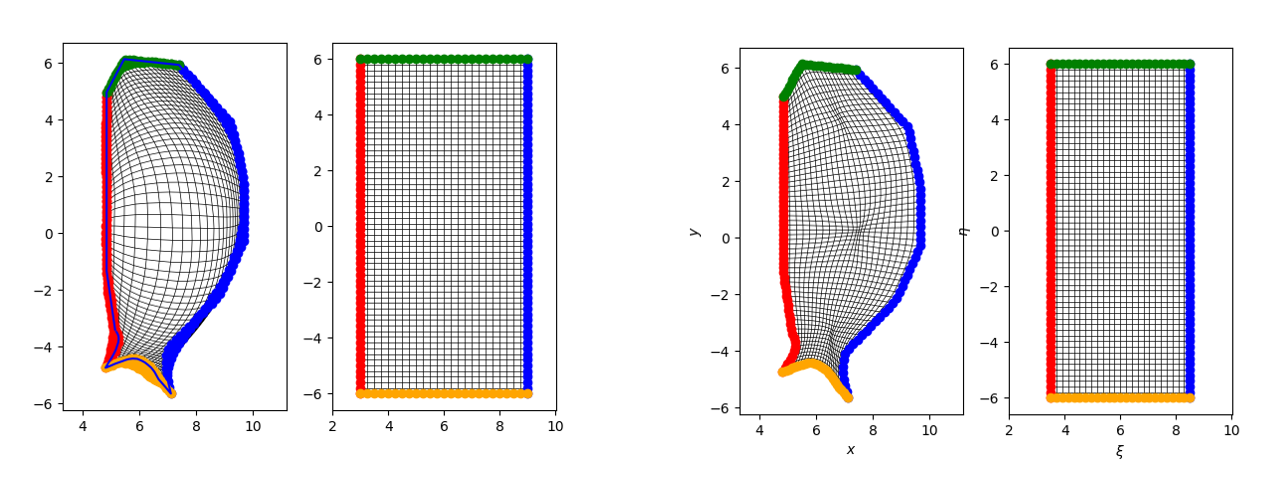}
    \caption{Sparser and denser grids around specific points.}
    \label{fig:denser_grids_around_points}
\end{figure}

We now present PINN-based grid generation method for sparser and denser grids around specific points. 
Figure \ref{fig:denser_grids_around_points} demonstrates these effects where the grids become sparser and denser at the regions around some points.

In the problem of grid generation, a mathematical framework for achieving the densification/sparsification of specific regions by adjusting the smooth term of the loss function can be constructed as follows:
\begin{equation}
L_{eq}=\frac{1}{N_f}\sum_{i = 1}^{N_f}((\nabla^2 x - P(\xi,\eta))^2 +(\nabla^2 y- Q(\xi,\eta)) ^2) , \quad \forall (\xi,\eta) \in \Omega_c
\label{eq:smooth_eq_PQ}
\end{equation}
where

\begin{equation}
    \nabla^2 x = \frac{\partial^2 x}{\partial \xi^2} + \frac{\partial^2 x}{\partial \eta^2} , 
\end{equation}
\begin{equation}
 \nabla^2 y = \frac{\partial^2 y}{\partial \xi^2} + \frac{\partial^2 y}{\partial \eta^2},
\end{equation}
and the expressions of $P(\xi,\eta)$ and $Q(\xi,\eta)$ are

\begin{equation}
P = \alpha \cdot \frac{\xi-\xi_0}{|\xi-\xi_0|} \cdot e^{-\beta\left[(\xi-\xi_0)^2 + (\eta-\eta_0)^2\right]} ,
\end{equation}

\begin{equation}
Q = \gamma \cdot \frac{\eta-\eta_0}{|\eta-\eta_0|} \cdot e^{-\beta\left[(\xi-\xi_0)^2 + (\eta-\eta_0)^2\right]},
\end{equation}
and ($\xi_0$,$\eta_0$) is the coordinate of the densified (sparsified) point in the computational space. By changing the signs of the functions $P(\xi,\eta)$ and $Q(\xi,\eta)$, the conversion function between sparsification and densification can be achieved. The parameters, $\alpha$,$\beta$ and $\gamma$ are parameters that control the degree of distortions.

\subsection{Mesh merging}

When solving physical problems, if different types of grids need to be generated for the interior and exterior of a device, PINN can be used to generate the structured grids separately, followed by grid merging process. In this way, the originally complex physical problem can be transformed into a computational task on two adjacent rectangular regions, thereby effectively reducing the computational complexity and improving the computational efficiency and accuracy.

\begin{figure}
    \centering
    \includegraphics[width=\linewidth]{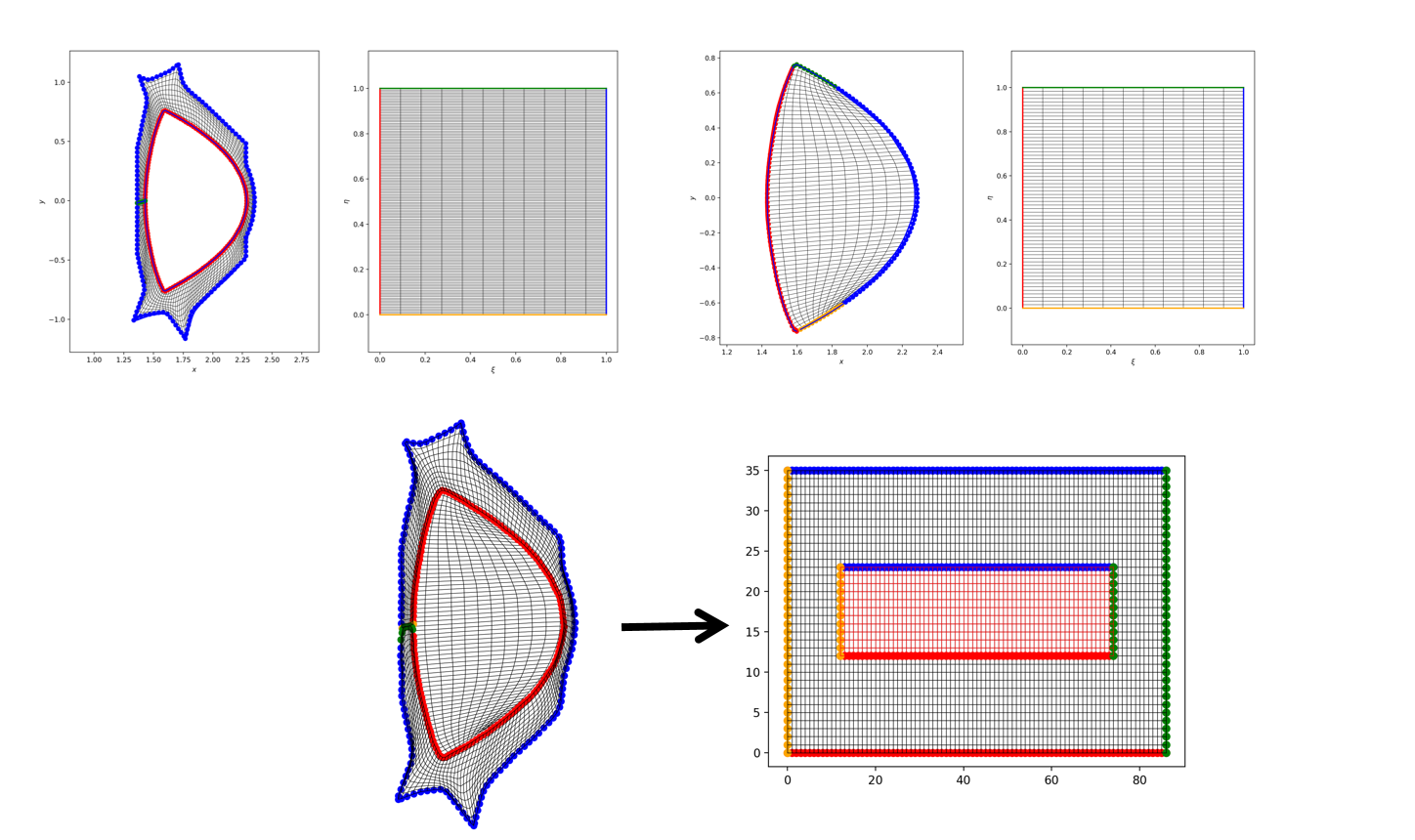}
    \caption{Grid merging}
    \label{fig:Mesh Merging}
\end{figure}

As shown in Fig. \ref{fig:Mesh Merging},  meshes can be merged to a unified one. The method of  mesh merging  is  straightforward by taking the common sampling boundary points  in physical space and by surrounding the rectangular grids of external region around the grids of internal region in computational space.  
PDEs  can be solved  on the irregular  external region and the internal region separately, or they can be solved on the unified region.  This method is more favorable for occasions where physical equations are different in different regions.

\section{Conclusion}

In this paper, the method of PINN is used to automatically generate structured grids and obtain the mapping relationship between irregular domains and regular domains, so that physical problems in irregular domains can be transformed and solved in rectangular domains. According to the characteristics of PINN, adding a loss function to the loss term can generate grids that meet special requirements. The grids generated by PINN can be used not only for traditional methods of solving PDE equations, such as FDM and the spectral method, but also for the AI based method, such as PICNN.



\begin{thebibliography}{10}
\expandafter\ifx\csname url\endcsname\relax
  \def\url#1{\texttt{#1}}\fi
\expandafter\ifx\csname urlprefix\endcsname\relax\def\urlprefix{URL }\fi
\expandafter\ifx\csname href\endcsname\relax
  \def\href#1#2{#2} \def\path#1{#1}\fi

\bibitem{astuto2024nodal}
C.~Astuto, D.~Boffi, G.~Russo, U.~Zerbinati, A nodal ghost method based on variational formulation and regular square grid for elliptic problems on arbitrary domains in two space dimensions, arXiv preprint arXiv:2402.04048 (2024).

\bibitem{HOLLBACHER2019186}
S.~Höllbacher, G.~Wittum, Rotational test spaces for a fully-implicit fvm and fem for the dns of fluid-particle interaction, Journal of Computational Physics 393 (2019) 186--213.
\newblock \href {https://doi.org/https://doi.org/10.1016/j.jcp.2019.05.004} {\path{doi:https://doi.org/10.1016/j.jcp.2019.05.004}}.

\bibitem{RAISSI2019686}
M.~Raissi, P.~Perdikaris, G.~Karniadakis, Physics-informed neural networks: A deep learning framework for solving forward and inverse problems involving nonlinear partial differential equations, Journal of Computational Physics 378 (2019) 686--707.

\bibitem{KHARA2024103709}
B.~Khara, E.~Herron, A.~Balu, D.~Gamdha, C.-H. Yang, K.~Saurabh, A.~Jignasu, Z.~Jiang, S.~Sarkar, C.~Hegde, B.~Ganapathysubramanian, A.~Krishnamurthy, Neural pde solvers for irregular domains, Computer-Aided Design 172 (2024) 103709.

\bibitem{Peskin_2002}
C.~S. Peskin, The immersed boundary method, Acta Numerica 11 (2002) 479–517.
\newblock \href {https://doi.org/10.1017/S0962492902000077} {\path{doi:10.1017/S0962492902000077}}.

\bibitem{SUKUMAR2022114333}
N.~Sukumar, A.~Srivastava, Exact imposition of boundary conditions with distance functions in physics-informed deep neural networks, Computer Methods in Applied Mechanics and Engineering 389 (2022) 114333.

\bibitem{McFall20091221}
K.~S. McFall, J.~R. Mahan, Artificial neural network method for solution of boundary value problems with exact satisfaction of arbitrary boundary conditions, IEEE Transactions on Neural Networks 20~(8) (2009) 1221 – 1233.
\newblock \href {https://doi.org/10.1109/TNN.2009.2020735} {\path{doi:10.1109/TNN.2009.2020735}}.

\bibitem{SHENG2021110085}
H.~Sheng, C.~Yang, Pfnn: A penalty-free neural network method for solving a class of second-order boundary-value problems on complex geometries, Journal of Computational Physics 428 (2021) 110085.
\newblock \href {https://doi.org/https://doi.org/10.1016/j.jcp.2020.110085} {\path{doi:https://doi.org/10.1016/j.jcp.2020.110085}}.

\bibitem{lu2021learning}
L.~Lu, P.~Jin, G.~Pang, Z.~Zhang, G.~E. Karniadakis, Learning nonlinear operators via {DeepONet} based on the universal approximation theorem of operators, Nature Machine Intelligence 3~(3) (2021) 218--229.

\bibitem{GAO2021110079}
H.~Gao, L.~Sun, J.-X. Wang, Phygeonet: Physics-informed geometry-adaptive convolutional neural networks for solving parameterized steady-state pdes on irregular domain, Journal of Computational Physics 428 (2021) 110079.

\bibitem{anderson1995computational}
J.~D. Anderson, Computational Fluid Dymanics :The Basic with Applications, McGraw-Hill Press, 1995.

\end{thebibliography}

\end{document}